\documentclass[fleqn,usenatbib]{mnras}

\usepackage{newtxtext,newtxmath}

\usepackage[T1]{fontenc}

\DeclareRobustCommand{\VAN}[3]{#2}
\let\VANthebibliography\thebibliography
\def\thebibliography{\DeclareRobustCommand{\VAN}[3]{##3}\VANthebibliography}

\usepackage{graphicx}	
\usepackage{amsmath}	

\usepackage[nolist]{acronym}

\newcommand{\TOTALNUMBERCHZEROPEAKFLUX}{339,128}

\begin{acronym}[AWGN]
\acro{ASKAP}{Australian Square Kilometre Array Pathfinder}
\acro{BANE}{Background and Noise Estimator}
\acro{FOV}{field of view}
\acro{FWHM}{Full Width at Half Maximum}
\acro{GPS}{Gigahertz-Peaked Spectrum}
\acro{MC}{Magellanic Cloud}
\acro{MWA}{Murchison Widefield Array}
\acro{LMC}{Large Magellanic Cloud}
\acro{RMS}{root mean squared}
\acro{SARAO}{South African Radio Astronomy Observatory}
\acro{SMC}{Small Magellanic Cloud}
\end{acronym}

\newcommand{\aegean}{\textsc{Aegean}}
\newcommand{\HII}{{H\,{\sc ii}}\xspace}
\def\HII{\hbox{H{\sc ii}}}

\title[MeerKAT LMC catalogue]{The MeerKAT 1.3\,GHz survey of the Large Magellanic Cloud: Point Source Catalogue}

\author[N. Rajabpour et al.]{
N. Rajabpour,$^{1}$\thanks{E-mail:}
M. D. Filipovi\'c,$^{1}$
W. D. Cotton,$^{2,3}$
Z. J. Smeaton,$^{1}$
A. C. Bradley,$^{1}$
E. J. Crawford,$^{1}$
\newauthor{
M. Ghavam,$^{1}$
O. K. Khattab,$^{1}$
and J. Th. van Loon$^{4}$
}
\\
$^{1}$Western Sydney University, Locked Bag 1797, Penrith South DC, NSW 2751, Australia\\
$^{2}$National Radio Astronomy Observatory, 520 Edgemont Road, Charlottesville, VA 22903, USA\\
$^{3}$South African Radio Astronomy Observatory (SARAO), 2 Fir Street, Black River Park, Observatory, Cape Town 7925, South Africa\\
$^{4}$Lennard-Jones Laboratories, Keele University, ST5 5BG, UK
}

\date{Accepted XXX. Received YYY; in original form ZZZ}

\pubyear{\the\year{}}

\begin{document}
\label{firstpage}
\pagerange{\pageref{firstpage}--\pageref{lastpage}}
\maketitle

\begin{abstract}
We present a point source catalogue of the \ac{LMC} using observations from the MeerKAT radio telescope across a broad frequency band centred at 1295\,MHz (L-Band) with 12 sub-bands and a resolution of 8\arcsec$\times$8\arcsec. 
We investigate source flux density distributions and spectral indices and compare these with previous similar radio catalogues such as those conducted with the \ac{ASKAP} telescope at 888\,MHz. This new MeerKAT survey reaches a lower average noise level of 11\,$\mu$Jy\,beam$^{-1}$, compared with the previous \ac{ASKAP} levels of 58\,$\mu$Jy\,beam$^{-1}$. This higher sensitivity and resolution allow the detection of a significantly larger number of faint sources. The catalogue includes 339,128 point sources; a substantial increase over the 54,612 sources detected by \ac{ASKAP} providing critical insights into the population distribution, spectral properties, positional accuracy, and overall improvement in source recovery and survey depth. 
\end{abstract}




\section{Introduction}

\acresetall

The nearby \ac{MC} system, including the \ac{LMC} and \ac{SMC}, is a nearby galaxy system, located in a relatively clear area of the sky. Thus, this system represents an ideal target for studying entire galaxies and galaxy populations, as it is nearby enough to be resolved in detail and is relatively unimpeded by Galactic extinction. In the last half-century, numerous radio surveys have studied the nearby \ac{MC} system, including the \ac{LMC} and \ac{SMC}. These surveys have significantly advanced our understanding of the various populations of sources and the overall galaxy distribution and evolution~\citep{1976MNRAS.174..393C, 1981MNRAS.194..693L, 1985PASA....6...72M, 1993AJ....105.1666G, filipovic1995radio, Filipovic1996, 1996AAS...188.6505W, 1998A&AS..130..421F, 1998PASA...15..128F, 1998A&AS..130..441F, Filipovic1998, 1998ApJ...503..674K, 2003ApJS..148..473K, 2003MNRAS.342.1117M, 2005AJ....129..790D, 2007MNRAS.382..543H, 2010MNRAS.402.2403M, 2018MNRAS.480.2743F, 2021MNRAS.506.3540P}. 

The recent advent of several new radio telescopes has been of paramount importance in studying the \ac{MC}. These telescopes, such as \ac{ASKAP}, MeerKAT, and \ac{MWA}, have excellent surveying capabilities~\citep{2009ASPC..407..446J,2021PASA...38....9H, 2016mks..confE...1J, 2013PASA...30....7T} and have studied the \acp{MC} with better resolution and sensitivity than previously possible. These recent surveys~\citep{2021MNRAS.506.3540P, 2019MNRAS.490.1202J, 2024MNRAS.529.2443C, 2018MNRAS.480.2743F} have allowed us an excellent view into the populations of these galaxies and advanced our understanding of the \acp{MC} and galaxy evolution as a whole.


The survey of the \ac{LMC} conducted with MeerKAT represents a notable advancement over previous radio-continuum investigations. MeerKAT's excellent resolution and sensitivity have been previously validated in other research~\citep{2022ApJ...925..165H, 2024MNRAS.529.2443C, 2024MNRAS.531..649G}. This paper presents a catalogue of point sources within the \ac{LMC} area derived from this MeerKAT survey of the \ac{LMC}. Further details of the survey itself can be found in  \citet{CottonLMC}. The data were obtained from images captured at a broadband frequency of 1295\,MHz ($\lambda$\,=\,23\,cm; L-Band) and 12 sub-bands. 

The structure of the paper is as follows: Section~\ref{sec:data} outlines the MeerKAT observations and the data used, Section~\ref{subsec:detection} discusses the source detection methodologies, the calculated spectral indices of the sources are discussed in Section~\ref{subsec:spectral indices}, and our final catalogue is presented in Section~\ref{subsec:catalogues}. In Section~\ref{sec:comparison} we compare our findings with previous point source catalogues in the direction of the \ac{LMC}, and in Section~\ref{sec:summary and conclusion} we summarise our conclusions.

\section{Data}
\label{sec:data}

We used observational data from MeerKAT; a summary is provided here, with further technical details provided in \citet{CottonLMC}. The \ac{LMC} was observed using the L-Band frequency range (856 to 1712\,MHz) of the MeerKAT Array~\citep{Jonas2016, 2018ApJ...856..180C, DEEP2}, between 24$^{\rm th}$ August 2019 to 18$^{\rm th}$ November 2019 (project code: SSV-20180505-FC-02), using 8 second integrations and 4096 spectral channels across the band. All four combinations of the two linearly polarised feeds were recorded. The total observing time was 258.4 hours. 

Since the \ac{LMC} is substantially larger than the MeerKAT antenna beam, it was covered by a mosaic of 207 pointings in a hexagonal grid centred on RA\,=\,05$^{\rm h}$23$^{\rm m}$34.0$^{\rm s}$, Dec\,=\,$-$69$^{\circ}$45\arcmin22\farcs0. The offset between pointing centres is 29\farcm6, giving relatively uniform sensitivity. 

Calibration was performed using the calibrator PKS~0408$-$65 for flux density, delay, bandpass, and astrometric solutions, and J0521+1638 (3C138) was used for polarisation solutions. Calibration and editing were interleaved, and the flux scale was referenced to PKS~B1934$-$638, following the process of \citet{Reynolds94}. 

Imaging was done using the \texttt{MFImage} task in the Obit package~\citep{2018ApJ...856...67C}, resulting in Stokes $I$, $Q$, $U$, and $V$ images for 5\% fractional bandwidth sub-bands. CLEANing thresholds were set to 60, 40, and 20\,$\mu$Jy\,beam$^{-1}$ for $I$, $Q/U$, and $V$ Stokes parameters respectively. Direction-independent self-calibration was applied to all pointings, with additional ``peeling'' applied where necessary to mitigate direction-dependent effects.

Astrometric corrections were derived using the MilliQuas source catalogue and applied during mosaicing. The final mosaic images (six images with 5$^\circ\times5^\circ$ \ac{FOV}) were constructed using a weighted linear combination of overlapping pointings, which were convolved to a uniform 8~arcsec \ac{FWHM} resolution. Individual sub-band images were formed into mosaics, and the broadband mosaic images were derived from the sub-band mosaics.

The final data products, including visibility data and image mosaics, are available on the \ac{SARAO}\footnote{\url{https://archive.sarao.ac.za}} archive in various formats, allowing further spectral and polarisation analysis as required.

\begin{figure*}
    \centering
	\includegraphics[width=\textwidth]{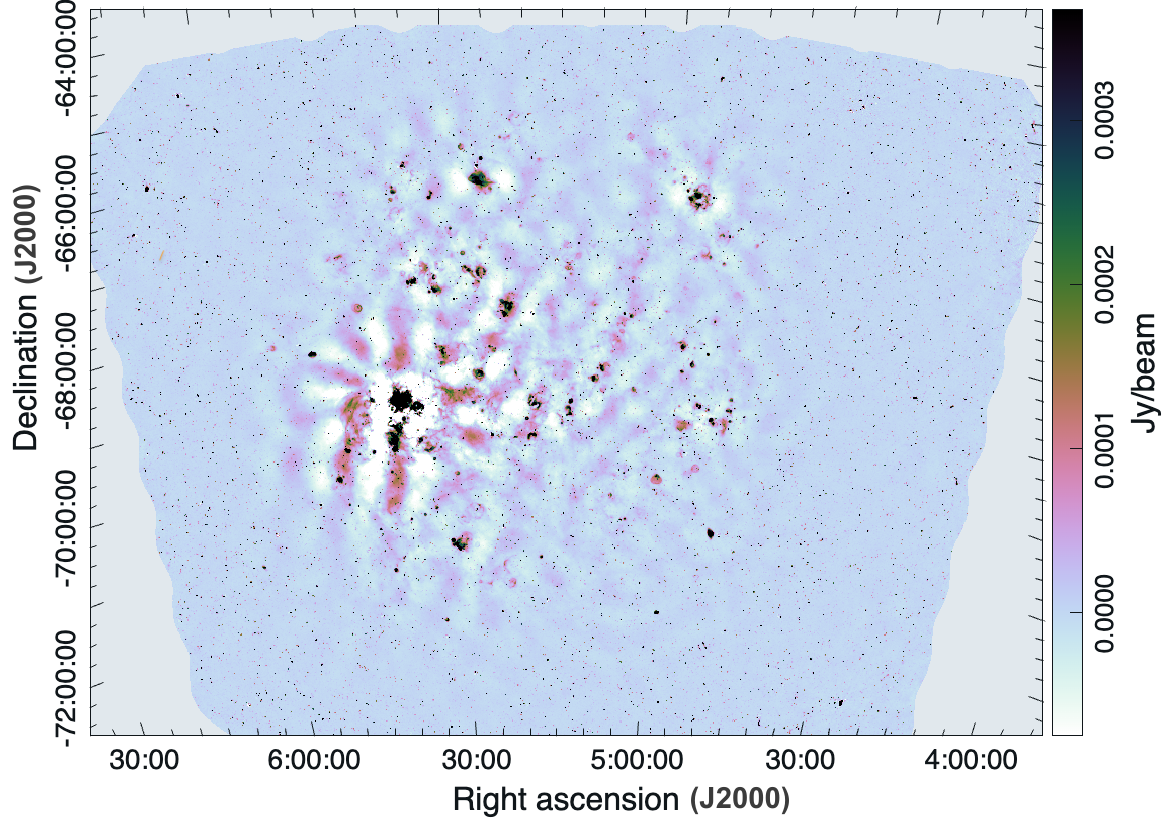}
    \caption{Stitched image of the \ac{LMC} created from six overlapping $5^\circ \times 5^\circ$ fields, created as described in text and \citet{CottonLMC}. The image is linearly scaled.}
    \label{fig:LMCWHOLE}
\end{figure*}

The six $5^\circ\times5^\circ$ mosaics were combined into a larger super-mosaic (see Fig.~\ref{fig:LMCWHOLE}) using the \texttt{SWarp} package~\citep{2002ASPC..281..228B}. This super-mosaic highlights several key features of the \ac{LMC}, such as the bright central bar region, an elongated concentration of ionised gas associated with \HII\ regions extending from the centre of the galaxy, and 30~Doradus (30~Dor), a massive star-forming complex located in the northeast. There are several artifacts present around the 30~Dor area, some of which are likely due to missing short-spacing information present in the MeerKAT data. This is a consequence of radio interferometric images and can limit sensitivity to some large-scale emission. It produces artifacts such as sidelobes, negative bowls, and distortions around bright extended sources such as 30~Dor. More detailed information about the imaging and imaging artifacts can be found in \citet{CottonLMC}. 

This larger mosaic provides an overview of the entire \ac{LMC} and enables us to more efficiently search for point sources. We detail the point source searching methodology and catalogue generation below, and we have followed a process similar to that of \citet{2024MNRAS.529.2443C} for the \ac{SMC}.


\section{MeerKAT LMC Point Source Catalogue}
\label{sec:source catalogue}

\subsection{Source Detection}
\label{subsec:detection}

Following a similar approach to previous large-scale surveys~\citep[e.g.;][]{2024MNRAS.529.2443C, 2021MNRAS.506.3540P, 2019MNRAS.490.1202J}, we utilised the \aegean\ software, an algorithm designed for compact source detection and flux extraction~\citep{hancock2018}. This tool was used to generate a point source catalogue from the broadband image (also referred to as channel-0 throughout) and to extract flux densities from the channel-0 image. This generated catalogue was then used as the base catalogue to search all sub-band images and extract the sub-band flux density if a point source is found. The results are detailed in Table~\ref{tab:subchannumbers}. We note that the sub-band images for channels 8 and 9 are not included in the analysis, as these channels were blanked due to significant radio frequency interference.


\begin{table}
    \caption{Source detection results across frequency channels. The table lists the frequency (in MHz) for each channel and the total number of point sources detected per channel. Channels 1, 8, and 9 are not included as discussed in the text.}
    \centering
    \begin{tabular}{ccc}
    \hline
    Channel & Frequency (MHz)
    & Total Sources Found\\
    \hline
    0 & 1295.0 & 
    \TOTALNUMBERCHZEROPEAKFLUX  \\
    2 & 908.0 & 
    144,532\\
    3 & 952.3 & 
    96,179\\
    4 & 996.6 & 
    158,543\\
    5 & 1043.4 & 
    160,094\\
    6 & 1092.8 & 
    148,470\\
    7 & 1144.6 & 
    108,360\\
    10 & 1317.2 & 
    135,544\\
    11 & 1381.2 & 
    153,188\\
    12 & 1448.1 & 
    149,762\\
    13 & 1519.9 & 
    97,817\\
    14 & 1593.9 & 
    49,286\\
    15 & 1656.2 & 
    93,838\\
    \hline
    \end{tabular}
    \label{tab:subchannumbers}
\end{table}

\subsubsection{Broadband image}
\label{subsec:imagemain}

\begin{figure*}
    \centering
    \includegraphics[trim={0 0 0 0},width=1\textwidth]{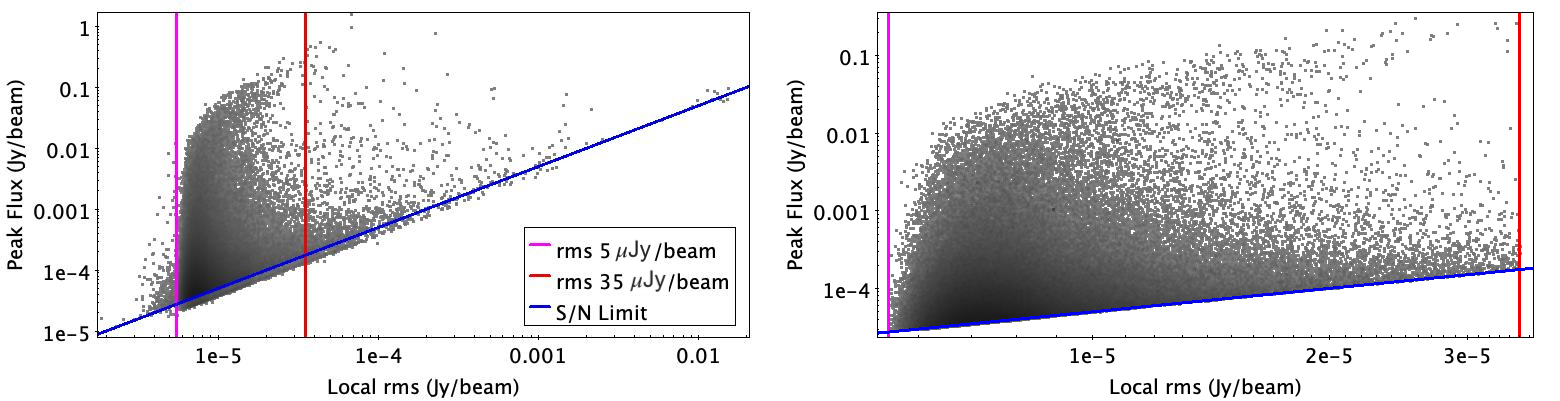} 
    \caption{
    The plot of the relationship between local \ac{RMS} noise levels and the peak flux density of sources from the initial catalogue (left) and the final catalogue (right). The distribution is generally smooth, with noticeable density excesses in regions where \ac{RMS} exceeds 35~$\mu$Jy~beam$^{-1}$
  (marked by a vertical red line) and where \ac{RMS} is below 5~$\mu$Jy~beam$^{-1}$ 
  (marked by a vertical magenta line). Detections below the applied signal-to-noise threshold (blue line) were removed from the final catalogue.}
    \label{fig:rms_peak}
\end{figure*}

The initial catalogue was generated from the broadband image using the \aegean\ software. The steps are outlined as follows: 

\begin{itemize}
    \item \textbf{Background estimation:} First, we estimated background noise levels for the broadband image using the (\textsc{\ac{BANE}}) tool \citep{hancock2018} found in the \aegean\ software package. This tool measures the background noise levels in an image and generates a background \ac{RMS} noise map for input into \aegean. 
    \item \textbf{Finding sources:} The broadband image and the generated background noise image are input into \aegean\ which runs a source finding algorithm to detect point sources. The search was conducted using the default \aegean\ parameters~\citep{hancock2018}, with the parameters {\it seed clip}  5$\sigma$ and the {\it flood clip}  4$\sigma$. The sources are fit with Gaussian profiles. The initial catalogue contains 386,520 sources.
    \item \textbf{Analyse Source Distribution:} We then plot the generated catalogue on the broadband image to analyse the source distribution and search for any major anomalies. We find that the sources are concentrated near the central super-mosaic, as expected due to the higher sensitivity in this area, and the density decreases towards the edge of the image due to lower image sensitivity. 
    \item \textbf{Image and Noise \ac{RMS}:} The \ac{RMS} image produced during the \ac{BANE} process was visually examined. We find that the noise levels increase rapidly towards the edges of the image due to reduced sensitivity and primary beam attenuation. We also note that there were higher \ac{RMS} noise levels surrounding the 30~Dor region, due to its complicated nature, but find no other anomalies for the rest of the image.
    \item \textbf{Local RMS vs Peak Flux Ratio:} We plot the relationship between the local \ac{RMS} values and the peak flux densities from the generated \aegean\ catalogue (see Fig.~\ref{fig:rms_peak}, left). The distribution of the sources appears generally smooth, but some increased scatter is observed at both the lowest and highest \ac{RMS} noise levels. We thus investigate the nature of these sources and perform two \ac{RMS} noise level cuts on the data.
    \item \textbf{Upper Noise Level Cut:} From the distribution shown in Fig.~\ref{fig:rms_peak} (left), we determine two areas to perform an upper and lower cut on the data. We find a distinct decrease in source density at \ac{RMS} levels above $\sim$35\,$\mu$Jy\,beam$^{-1}$ (red vertical line in Fig.~\ref{fig:rms_peak}, left). The sources above this level have higher peak flux densities, due to only the brightest sources being detected above the local \ac{RMS} noise level. These areas of higher \ac{RMS} noise most likely represent complicated environmental areas, such as areas of diffuse extended emission, or areas with significant imaging artifacts present. These would include areas around bright sources, such as 30~Dor, or areas near the image edges. Thus, these points are excluded from the final dataset.
    \item \textbf{Lower Noise Level Cut:}
    We also note a specific region of interest where the \ac{RMS} values drop below 5\,$\mu$Jy\,beam$^{-1}$, (magenta vertical line in Fig.~\ref{fig:rms_peak}, left). The sources below this limit deviate from the smooth distribution seen at the higher \ac{RMS} levels and so were manually checked. We conducted a manual examination of $\sim$15 per cent of these sources and determined that most were spurious noise artifacts. We found a significant number of these were detected in or near bright sources, particularly in negative bowl areas. Therefore, these artifacts were creating artificially low \ac{RMS} noise levels and false positive detections by \aegean. Thus, these points are similarly excluded from the final dataset.
    \item \textbf{Signal to Noise Cut:} There are also a number of detections that appear as sources but actually fall below the established detection parameters of \aegean. These sources were mainly found to be artifacts, and thus to exclude them, we implement a manual cut, removing all sources with a signal-to-noise ratio less than 5 (blue diagonal line in Fig.~\ref{fig:rms_peak}, left). After these three cuts, the final distribution is shown in Fig.~\ref{fig:rms_peak} (right). The final catalogue has 339,128 point sources.
    \item \textbf{Flux Density Ratio:} In the resulting catalogue, we also found sources where the integrated flux density value was lower than the peak flux density value. This result can occur in source finding algorithms if the algorithm does not sufficiently resolve some sources properly. To account for this, we have replaced the integrated flux density value with the peak flux density value where this occurs, to ensure consistency in the flux density measurements.
\end{itemize}


\begin{figure*}		    
    \centering
	\includegraphics[width=\textwidth]{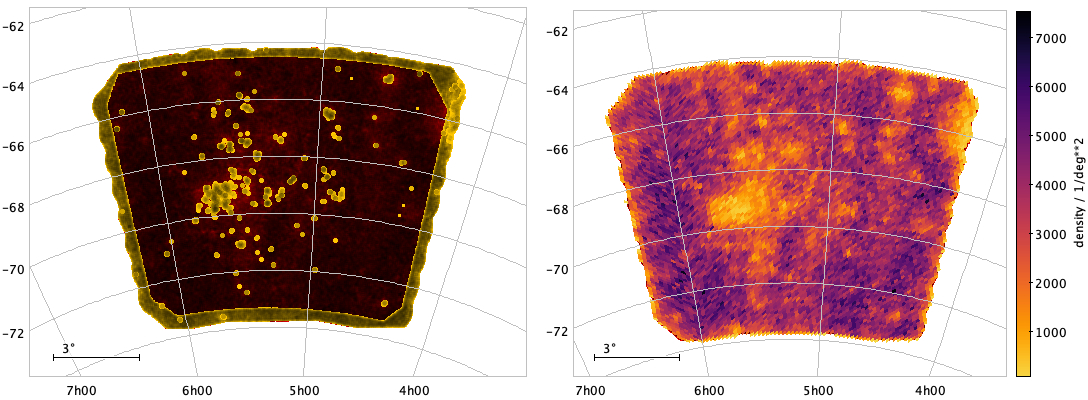}
	\caption{Source distribution maps of the MeerKAT LMC field. The green regions in the left figure correspond to sources located at the edge of the field, and the yellow regions in the left represent regions with an average \ac{RMS} exceeding 35~$\mu$Jy~beam$^{-1}$. The entire point source density of sources detected using \aegean\ are shown in the right image. Scale bars are shown in the bottom left corners of each image, and the images are linearly scaled.}
	\label{fig:distribution}
\end{figure*}	


\subsubsection{Sub-band images}
\label{subsubsec:subbandimages}

The final catalogue generated by running \aegean\ on the broadband image and performing the subsequent quality cuts has 339,128 sources. This catalogue was then used as input into \aegean\ along with the sub-band images to attempt to extract flux density measurements for these sources. The input catalogue was used to force-fitting of \aegean\ to the catalogue regions, ensuring that only the catalogued sources are measured. Some point sources were not detected in the sub-band images due to reduced sensitivity and higher noise levels in the narrower channels, and so these flux densities are not recorded. 

Using the default parameters, we found 178,559 sources were detected in the broadband image but had no corresponding detections in any of the sub-band images. This occurs because many faint sources are strong enough to be detected in the broadband image but too weak to be individually detected in the noisier, narrower sub-bands. Therefore, to improve the source measurements, we adjusted two of the default \aegean\ parameters: we set the {\it seed clip} parameter to be 4$\sigma$\ and the {\it flood clip} parameter to be 3$\sigma$. These values allowed the detection of fainter sources in the sub-band images, and as we are using the forced fitting algorithm, and thus only searching in regions where point sources are known to be located, this does not introduce significant erroneous measurements. This led to the detection of and additional 97,550 sources in the sub-band images, thus reducing the number of sources detected solely in the broadband image to 81,009.

\subsection{Spectral Indices}
\label{subsec:spectral indices}


\begin{figure*}
	\centering
	\includegraphics[width=\textwidth]{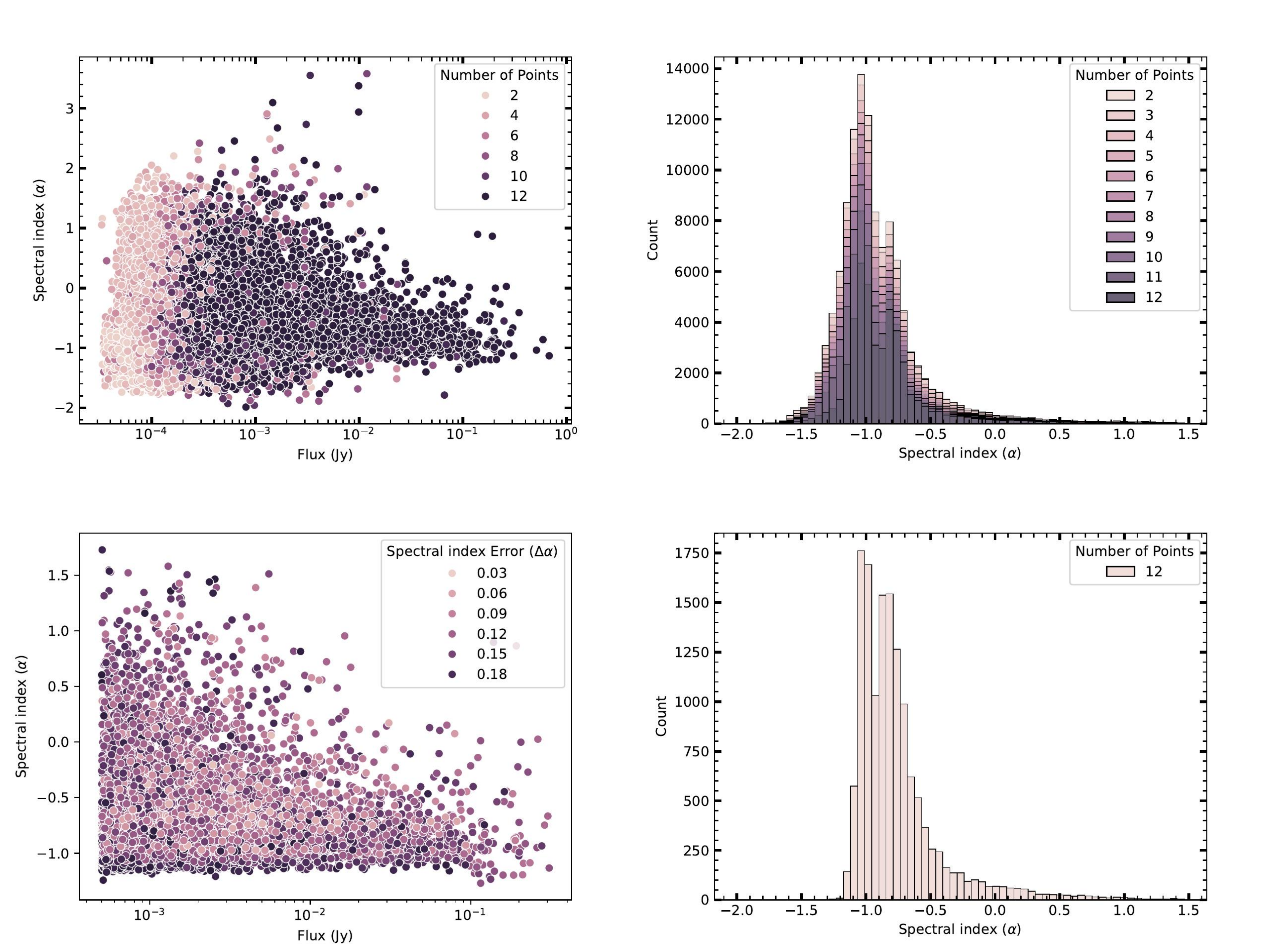}~
    \caption{
    Distribution of spectral indices across 12 different frequencies from MeerKAT. 
    {\bf Top Left}: Scatter plot showing the distribution of spectral indices per integrated flux density. The colours represent the number of channels used in the spectral index fit.
    {\bf Top Right}: Stacked histogram showing the point source spectral index distribution. The colours represent the number of channels used in the spectral index fit.  
    {\bf Bottom Left}: The distribution of spectral indices per integrated flux densities for 14,065 sources. This group of sources were restricted to those with only 12 frequency-flux density measurements as well as the flux density $>$0.5\,mJy, spectral index error $\Delta \alpha < 0.2$, and $\chi_\nu^2 < 1$. The colour bar represents the spectral index fit error. 
    {\bf Bottom Right}: The distribution of spectral indices for this sample of 14,065 sources.
    In all four figures, light colours represent small numbers, and dark colours represent high.
}
    \label{fig:ch5_check_ch_missing2}
\end{figure*}





For each source detected in at least two separate sub-band images, we calculate a spectral index $\alpha$, defined as $S\propto\nu^\alpha$~\citep{2021pma..book.....F}. The spectral index values are calculated for 200,163 sources. We also calculate the reduced $\chi^2$ value to analyse the goodness of the spectral index fit. As expected, we find that the reduced $\chi^2$ value improves as the number of sources (n) included in the fit increases. Since these flux density measurements are taken within an 800\,MHz band, small errors in sub-band flux density can lead to significant errors in the spectral index. Therefore, we restrict the analysis to only include sources with measurements from all 12 sub-bands. We also restrict this subset to only include sources with a broadband flux density $>0.5$\,mJy, reduced $\chi^2<$1, and $\Delta\alpha<0.2$. This leaves 14,065 sources for which we calculate the spectral index and use this to generate our population distribution (see Fig.~\ref{fig:ch5_check_ch_missing2} Bottom Right). We calculate a mean spectral index of $\alpha=-0.76$ (STDDEV\,=\,0.46), with a median of $\alpha=-0.88$ (see Figs.~\ref{fig:ch5_check_ch_missing2}).

Results show that our spectral index distribution aligns with previous studies, and the population displays a distinct ``long-tail'' of sources characterised by flat and inverted spectra, where the spectral index falls within the range of 0 to 1.5. Within this subset, we propose that sources exhibiting well-constrained spectral indices are predominantly \ac{GPS} sources or variable quasars, as supported by findings from \citet{2018MNRAS.477..578C}.

\subsection{Catalogues}
\label{subsec:catalogues}


\begin{table*}
    \begin{center}
        \caption{
        Example of detected sources in the catalogue, including their positions and flux density measurements at 1295 MHz. The columns provide the source name, right ascension (RA) and declination (Dec) in J2000 coordinates, peak flux density (S$_{\rm peak\,1295\,MHz}$) in mJy~beam$^{-1}$, integrated flux density (S$_{\rm int\,1295\,MHz}$) in mJy, spectral index ($\alpha$) with its uncertainty ($\Delta\alpha$), reduced chi-squared ($\chi^2_\nu$) for the spectral fit, and the number of data points (n) used in the spectral analysis.
        }
        \begin{tabular}{lllccccc}
        \hline
        Name          & RA (J2000)      &  Dec (J2000)  & S$_{\rm peak\,1295\,MHz}$     & S$_{\rm int\,1295\,MHz}$       &   $\alpha\pm\Delta\alpha$ & $\chi^2_\nu$ & n\\
              &      (h m s)           &    ($^\circ$\,\arcmin\ \arcsec)           &(mJy~beam$^{-1}$)& (mJy)         &  &  \\
        \hline
        J042423--663526 & 04:24:22.8 & --66:35:26 & 0.729$\pm$0.007 & 0.834$\pm$0.010 & --0.63$\pm$0.15& 7.99 & 12 \\
        J042424--665228 & 04:24:24.2 & --66:52:28 & 0.692$\pm$0.007 & 0.880$\pm$0.010 & --0.87$\pm$0.16 & 8.74 & 12 \\
        J042426--654453 & 04:24:25.9 & --65:44:53 & 0.914$\pm$0.007 & 1.664$\pm$0.016 & --0.91$\pm$0.12 & 8.60  & 12 \\
        J042428--655727 & 04:24:28.0 & --65:57:26 & 0.714$\pm$0.007 & 0.970$\pm$0.011 & --0.86$\pm$0.17 & 4.86 & 12 \\
        J042428--652450 & 04:24:28.1 & --65:24:49 & 0.502$\pm$0.008 & 0.771$\pm$0.010 & --0.75$\pm$0.18 &  4.29 & 12 \\
        J042429--721146 & 04:24:28.6 & --72:11:46 & 0.684$\pm$0.006 & 0.684$\pm$0.007 & --0.77$\pm$0.14 & 9.44 & 12 \\
        J042430--725338 & 04:24:29.8 & --72:53:38
        & 0.590$\pm$0.009 & 0.590$\pm$0.010 & --0.69$\pm$0.16 &  4.22 & 12 \\
       J042431--700125 & 04:24:30.8 & --70:01:25 & 0.542$\pm$0.006 & 0.551$\pm$0.007 & --0.96$\pm$0.16 & 8.27 & 12 \\
        J042432--724252 & 04:24:31.5 & --72:42:51 & 0.931$\pm$0.020 & 1.416$\pm$0.031 & --0.75$\pm$0.14 & 2.15 & 12 \\
        J042434--652102 & 04:24:34.1 & --65:21:01 & 0.637$\pm$0.007 & 0.641$\pm$0.009 & --0.74$\pm$0.17 &  6.67 & 12 \\
        \hline
        \label{tab:main}
        \end{tabular}
    \end{center}
\end{table*}

The final catalogue comprises 339,128 point sources, and an example of the final catalogue is given in Table~\ref{tab:main}. The final catalogue contains the source name (generated from the J2000 coordinates), the J2000 coordinates themselves, the broadband flux density for all sources, the flux density measurements for each sub-band it was detected in, the calculated spectral index value, the reduced $\chi^2$ value for the spectral index fit, and the number of points (n) used for the spectral index fitting.

\section{Comparison with previous catalogues}
\label{sec:comparison}

\subsection{Astrometry}
\label{subsec:astrometry}

To evaluate the precision of our catalogue's astrometry, we compare our results with those of two other catalogues.

\begin{figure*}
    \centering
\includegraphics[width=\columnwidth]{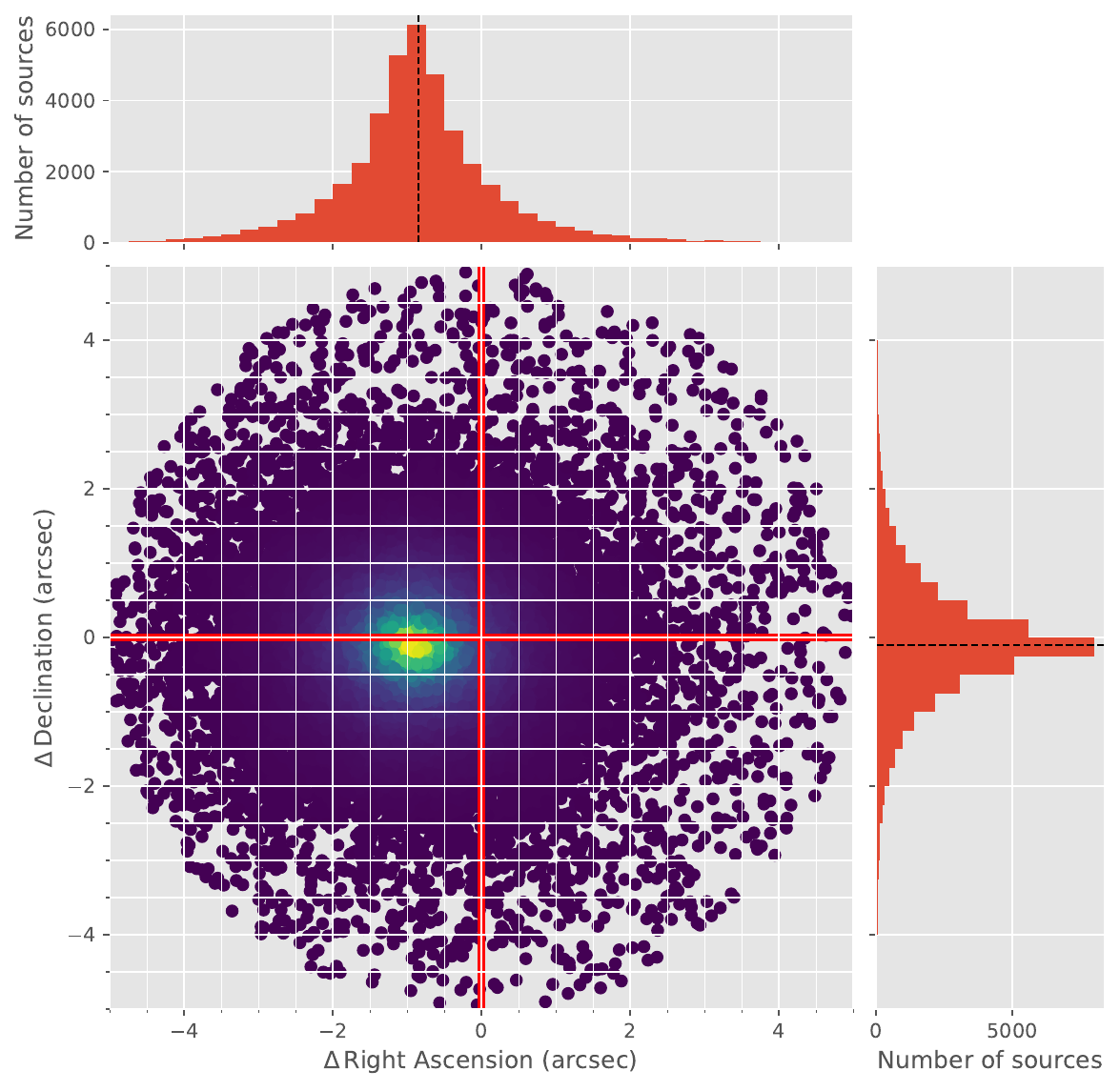}\includegraphics[width=\columnwidth]{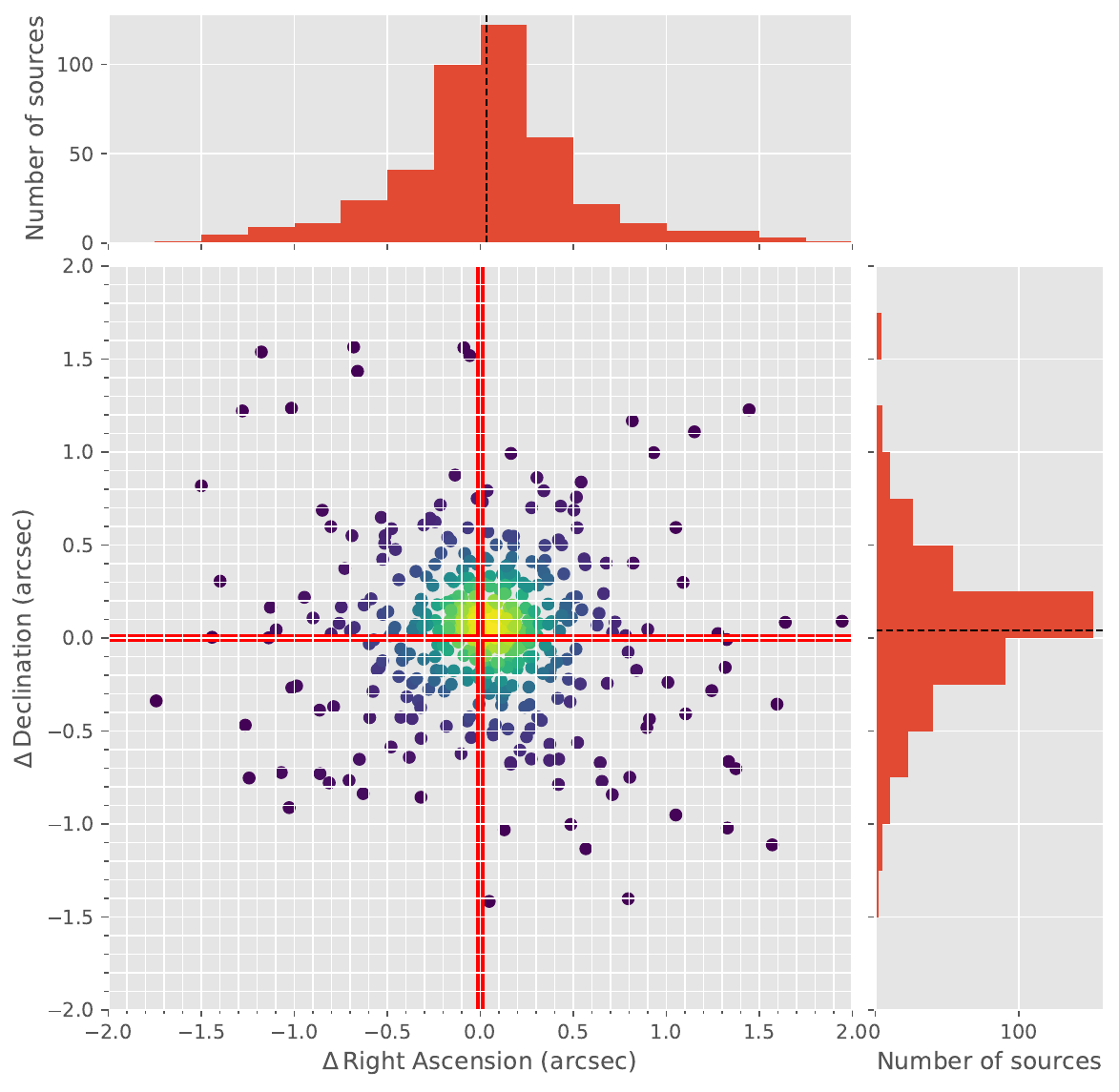}
            
    \caption{{\bf Left:} 
    Positional difference between the LMC MeerKAT and \ac{ASKAP} catalogue for 39,391 sources in common. The mean offsets are $\Delta$RA=$0.8$~arcsec (STDEV=1.09) and mean $\Delta$DEC=$0.1$~arcsec (STDEV=0.96).
    {\bf Right:}
 The positional differences between MilliQuas and MeerKAT catalogues were analysed for 423 sources identified through cross-matching both catalogues with a maximum separation threshold of 4~arcsec,
    with mean $\Delta$RA=--0.03~arcsec (STDEV=0.5) and mean $\Delta$DEC=--0.04~arcsec (STDEV=0.4).}
	\label{fig:position}
\end{figure*}

First, we compare our sources with those listed in the 888\, MHz \ac{ASKAP} catalogue of ~\citet{2021MNRAS.506.3540P}, which contains 54,612 sources. We crossmatched using the TOPCAT software~\citep{2005ASPC..347...29T}, using a 5~arcsec\ matching radius. This matching radius accounts for astrometric uncertainties between the observations and also for resolution differences between the two telescopes, $\sim$14~arcsec for \ac{ASKAP}~\citep{2021MNRAS.506.3540P} and 8~arcsec for MeerKAT. By cross-matching our source positions with those in the \ac{ASKAP} catalogue, we identify 39,391 sources in common. Consequently, 2,563 \ac{ASKAP} sources within the common field were not detected in the MeerKAT catalogue. This is likely due to differences in sensitivity, resolution, or possible source variability. 
Of the matched sources, we identified a positional offset of approximately $\Delta$RA\,=\,0.8~arcsec (STDDEV\,=\,1.09) and $\Delta$Dec\,=\,0.1~arcsec (STDDEV\,=\,0.96) (see Fig.~\ref{fig:position}, left). This level of discrepancy is deemed reasonable given that \ac{ASKAP} was still in its commissioning phase during the time of the observations~\citep{2021MNRAS.506.3540P}.

In addition to the \ac{ASKAP} comparison, we also compare our source positions against those of the MilliQuas catalogue~\citep{Milliquas}. The MilliQuas catalogue was originally used to correct for astrometric errors during the creation of the mosaic images, due to its good astrometrical precision \citep{CottonLMC}. We therefore expect no major deviations from the MilliQuas dataset, and this comparison is used to ensure that no astrometric errors were introduced during the source finding process with \aegean. For 423 sources that were matched within a 4~arcsec matching radius, we find mean offsets $\Delta$RA\,=\,$-$0.03~arcsec (STDDEV\,=\,0.5) and mean $\Delta$Dec\,=\,$-$0.04~arcsec (STDDEV\,=\,0.4) (see Fig.~\ref{fig:position}, right). These values demonstrate good astrometric agreement with the MilliQuas dataset, ensuring that no major astrometric issues were introduced by \aegean\ during the source finding process.

\subsection{Flux density}
\label{subsec:flux density}

\begin{figure*}		
			\includegraphics[width=0.65\textwidth]{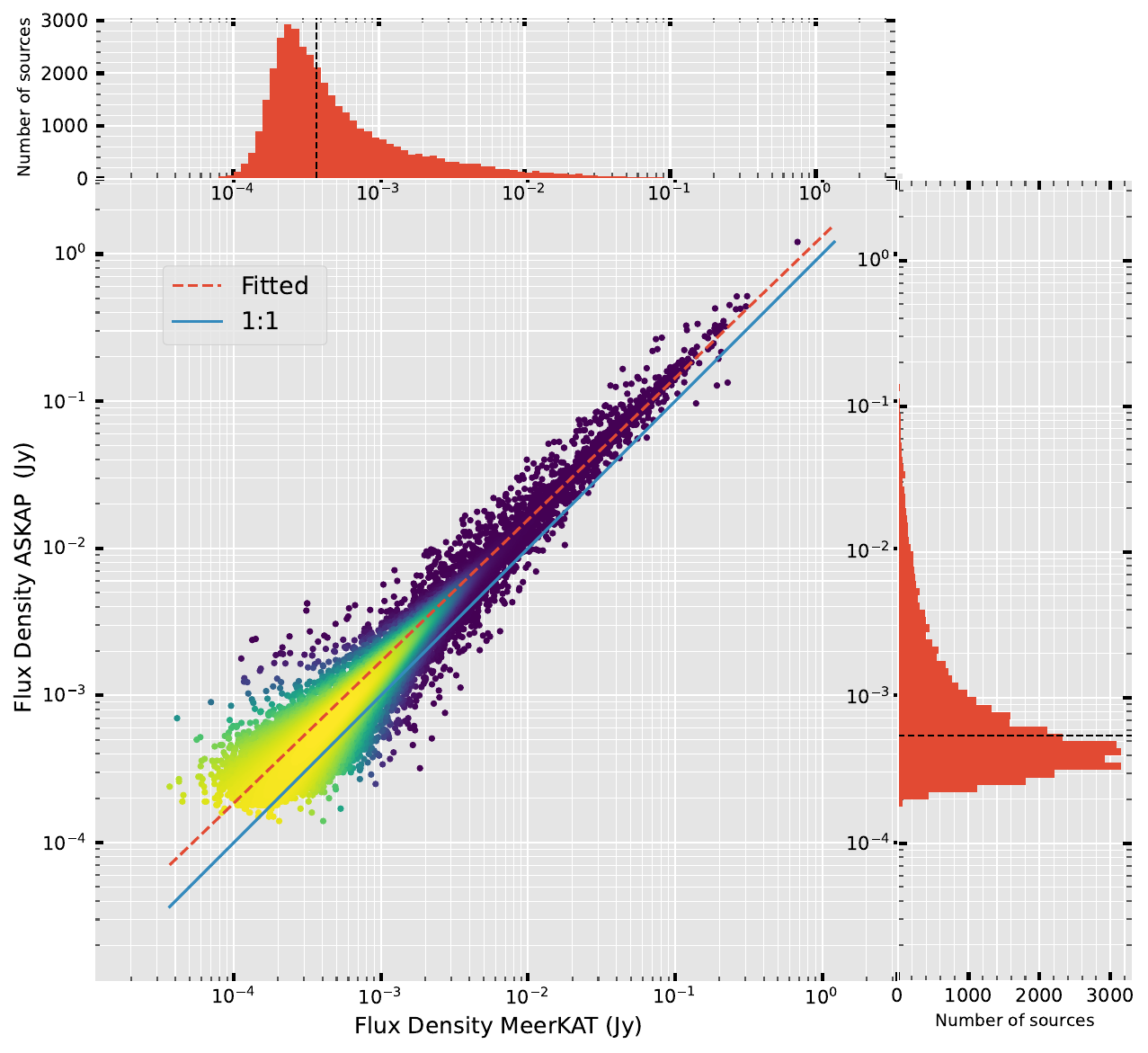}		\includegraphics[width=0.65\textwidth]{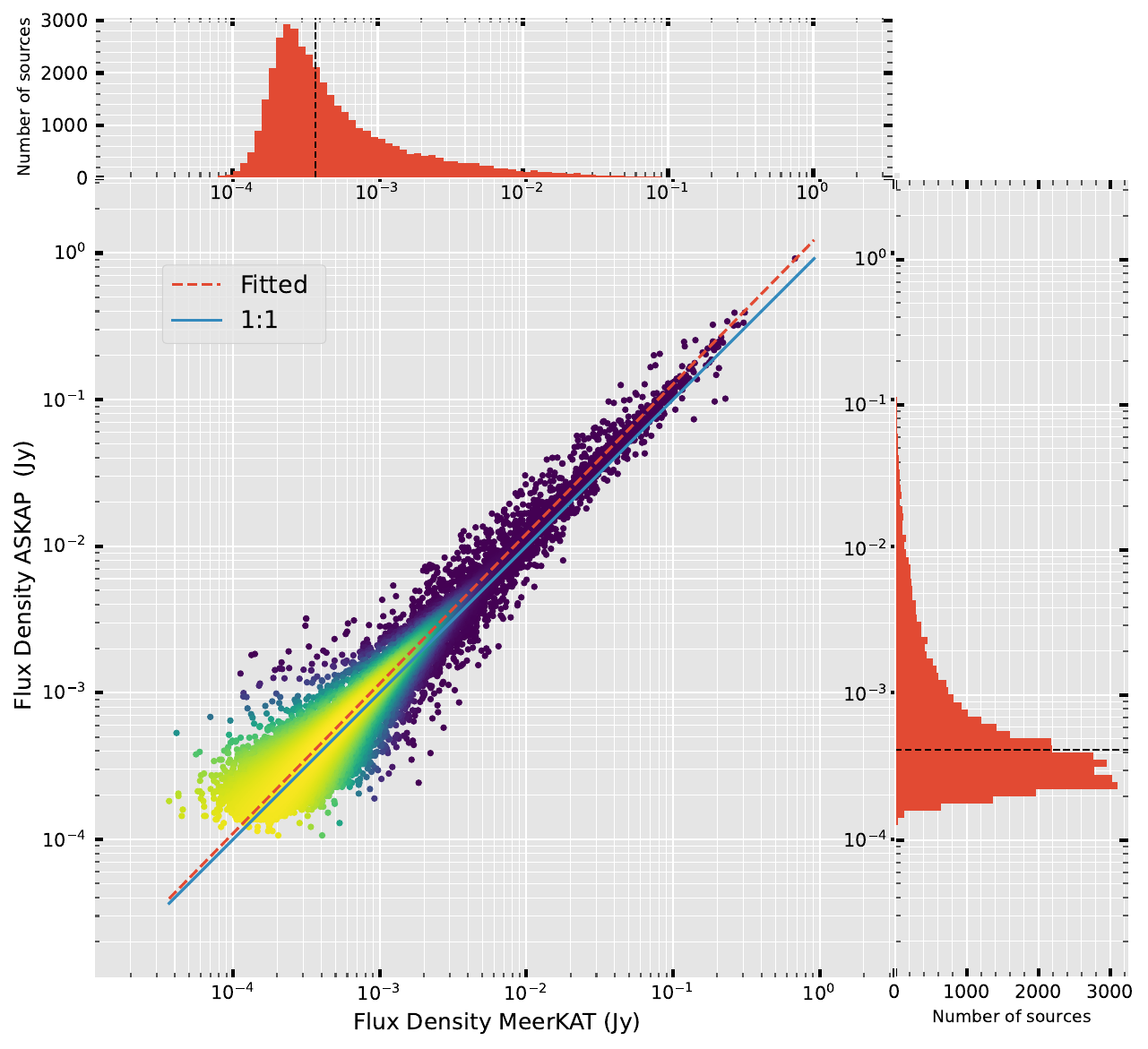}	
        \caption{
        The comparison of integrated flux density distributions includes 39,391 sources detected in both the 1295 MHz MeerKAT and 888 MHz \ac{ASKAP} observations \citep{2021MNRAS.506.3540P}. The blue line represents a 1:1 ratio line, and the red dashed line represents the linear best-fit slope for the data. The top and right-side histograms show the integrated flux density distributions for each survey, with the black line indicating the median integrated flux density. The data points are colour-coded to indicate local density, ranging from yellow (representing high density) to purple (indicating low density). MeerKAT fluxes have been scaled to 888\,MHz using a spectral index of $-0.76$, more details provided in the text. This scaling is to account for the offset caused by the different observing frequencies. The top panel shows the unscaled flux densities, and the bottom panel shows the flux densities once they have been scaled by $\alpha=-0.76$.}
		\label{fig:meerkat_vs_askap}
\end{figure*}

\begin{figure*}		
\includegraphics[width=\textwidth]{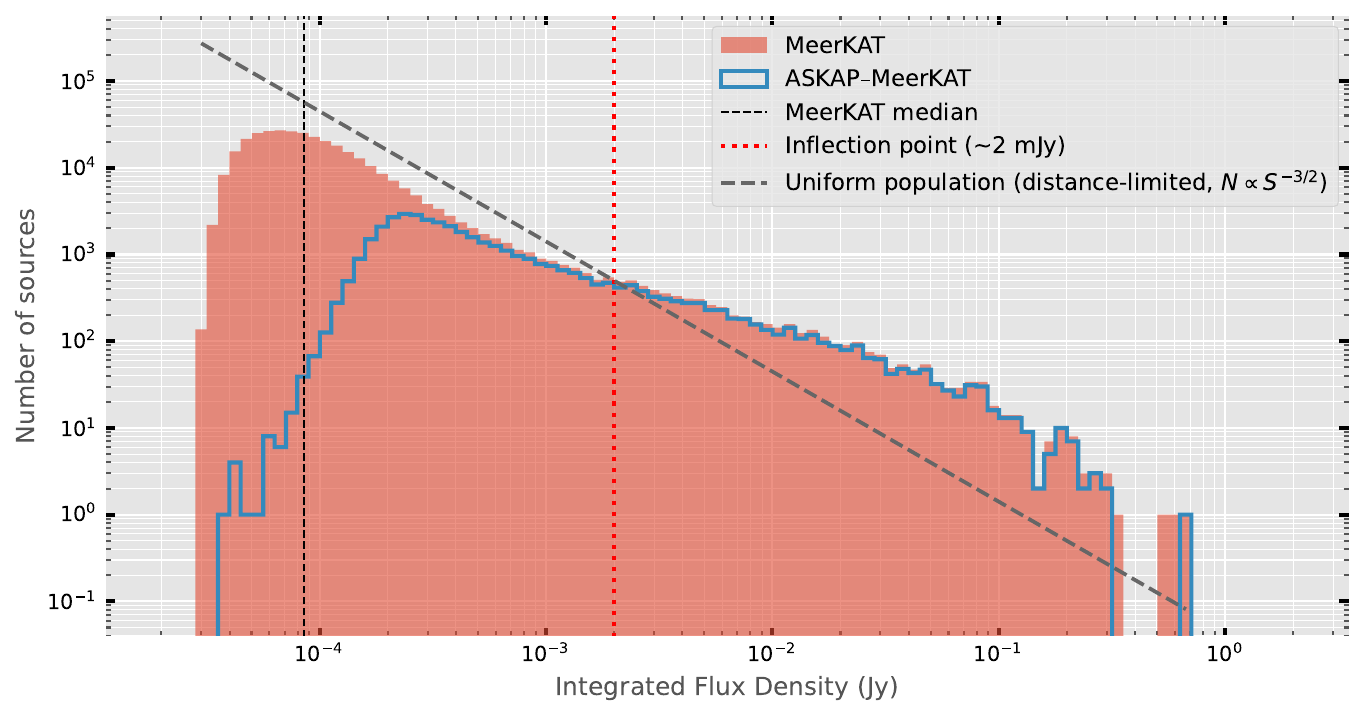}
        \caption{Histogram of integrated flux densities for MeerKAT sources (filled) and ASKAP--MeerKAT cross-matches (outline). The dashed red line indicates the inflection point at $\sim 2$~mJy. The grey dashed line shows the expected Euclidean scaling for a uniform, distance-limited population of identical-luminosity sources, for which \(N(>S) \propto S^{-3/2}\) (equivalently, \(N(>S)\,S^{3/2} = \mathrm{const}\)). Here, \(S\) is the flux density in Jy and \(N(>S)\) is the number of sources brighter than \(S\). The reference line is shown solely as a guide to the eye.}


		\label{fig:HISTOFLUX}
\end{figure*}

We also compared the integrated flux densities from \ac{ASKAP}~\citep{2021MNRAS.506.3540P} and MeerKAT (see Figs.~\ref{fig:meerkat_vs_askap} and \ref{fig:HISTOFLUX}) for the common set of 39,391 point sources present in both surveys. We show two lines in Fig.~\ref{fig:meerkat_vs_askap}; the blue line represents a 1:1 ratio line, and the dashed red line represents the linear best-fit slope for the data. The majority of point sources commonly display a steep spectral index of $\sim-0.8$, matching our measured mean spectral index of $\alpha\,=\,-0.76$ (STDDEV\,=\,0.46). This is demonstrated when the best-fit linear slope closely matches the 1:1 line, with an offset towards the higher \ac{ASKAP} values and no major discrepancies. This observed offset is partially due to differences in observing frequencies and the spread of the spectral index distribution. In Fig.~\ref{fig:ch5_check_ch_missing2} (Top left), we see an approximately normal distribution of spectral indices around the mean ($\alpha=-0.76$), and a long tail extending up to $\sim1-1.5$. This asymmetry in the distribution, also demonstrated by the standard deviation of 0.46, creates a spread in the data points for Fig.~\ref{fig:meerkat_vs_askap}, shown by the broadening of the distribution at lower flux densities. In Fig.~\ref{fig:meerkat_vs_askap} (bottom), the MeerKAT flux densities have been scaled to 888\,MHz using a spectral index of $-0.76$; that is, multiplied by $\left(\frac{1295}{888}\right)^{0.76} \approx 1.33$; to allow direct comparison with the \ac{ASKAP} fluxes. This scaling accounts for the offset between the best-fit linear slope and the 1:1 line observed in the left panel, which arises from differences in observing frequencies. After applying the spectral correction, the two lines exhibit close agreement.

Fig.~\ref{fig:HISTOFLUX} shows the distribution of integrated flux densities for the full MeerKAT catalogue and the ASKAP--MeerKAT cross-matched sources. The MeerKAT sample extends to substantially lower flux densities and contains a significantly larger number of faint sources. A clear change in slope is visible at approximately 2~mJy, marking an inflection point in the source counts. Below this level, the counts begin to deviate from the Euclidean expectation, likely reflecting incompleteness and sensitivity effects. The grey dashed line illustrates the idealised $N(>S) \propto S^{-3/2}$ scaling for a uniform, distance-limited population and is included for comparison.

\section{Summary and Conclusions}
\label{sec:summary and conclusion}

This study presents a comprehensive catalogue of 339,128 point sources detected toward the \ac{LMC} using the MeerKAT telescope at 1.3 GHz. The high sensitivity and resolution of the MeerKAT observations enabled the detection of faint sources and a detailed analysis of their properties, including flux densities, positions, and spectral indices.

The spectral index analysis highlighted a mean value of 
$\alpha$\,=\,--0.76 (STDDEV=0.46)
consistent with previous surveys, and identified a population of sources with flat spectra. The catalogue was cross-matched with the previous \ac{ASKAP} catalogue and was found to have only minor astrometric differences, likely attributed to the commissioning phase of \ac{ASKAP}.

\section*{Acknowledgements}

The MeerKAT telescope is operated by the South African Radio Astronomy Observatory, which is a facility of the National Research Foundation, an agency of the Department of Science and Innovation.
The National Radio Astronomy Observatory is a facility of the National Science Foundation, operated under a cooperative agreement by Associated Universities, Inc.

\section*{Data Availability}
The raw visibility data is available under project code SSV-20180505-FC-02 from the \ac{SARAO} archive (\url{https://archive-gw-1.kat.ac.za/public/repository/10.48479/zb9y-n292/index.html}) and the image data products can be obtained through \url{https://doi.org/10.48479/jrn4-ga52}. 




\bibliographystyle{mnras}
\bibliography{example} 





\bsp	
\label{lastpage}
\end{document}